\PassOptionsToPackage{table}{xcolor}
\documentclass[sigconf, screen, nonacm]{acmart}

\setcopyright{none}
\acmConference{}{}{}
\acmBooktitle{}
\acmDOI{}
\acmISBN{}


\usepackage{algorithm}
\usepackage{algorithmic}
\usepackage{textcomp}
\usepackage{xspace}
\usepackage{tabularx}
\usepackage{multirow}
\usepackage{pifont}
\usepackage[normalem]{ulem}
\usepackage{makecell}
\usepackage[most]{tcolorbox}
\usepackage{comment}

\usepackage{pifont}

\definecolor{rqbanner}{HTML}{F2E5B0}
\providecommand{\cmark}{\ding{51}}
\providecommand{\xmark}{\ding{55}}
\providecommand{\ourtool}{\textit{NEVA}\xspace}

\definecolor{insightbg}{HTML}{FBF7E9}

\usepackage{tikz}

\definecolor{figcalloutblue}{HTML}{0864C8}

\newcommand{\figcallout}[1]{%
  \tikz[baseline=-0.65ex]{%
    \node[
      circle,
      fill=figcalloutblue,
      text=white,
      inner sep=0pt,
      minimum size=10.5pt,
      font=\normalfont\fontsize{9}{10}\selectfont
    ] {#1};
  }%
}

\begin{document}
\renewcommand\footnotetextcopyrightpermission[1]{}

\title{NEVA: \underline{N}atural-Language Co-\underline{E}volution and \underline{V}alidation of Micro\underline{a}rchitectures}

\author{Khoi Le}
\authornote{Both authors contributed equally to this work.}
\email{khoile0315@tamu.edu}
\affiliation{%
  \institution{Texas A\&M University}
  \city{College Station}
  \state{Texas}
  \country{USA}
}

\author{Sayanti Jana}
\authornotemark[1]
\email{sayantijana008@tamu.edu}
\affiliation{%
  \institution{Texas A\&M University}
  \city{College Station}
  \state{Texas}
  \country{USA}
}

\author{Matthew DeLorenzo}
\email{matthewdelorenzo@tamu.edu}
\affiliation{%
  \institution{Texas A\&M University}
  \city{College Station}
  \state{Texas}
  \country{USA}
}

\author{Jeyavijayan (JV) Rajendran}
\email{jv.rajendran@tamu.edu}
\affiliation{%
  \institution{Texas A\&M University}
  \city{College Station}
  \state{Texas}
  \country{USA}
}

\author{Paul Gratz}
\email{pgratz@tamu.edu}
\affiliation{%
  \institution{Texas A\&M University}
  \city{College Station}
  \state{Texas}
  \country{USA}
}

\begin{abstract}

Recent large language model (LLM)-guided evolutionary approaches have demonstrated the potential to automate microarchitectural design exploration. However, existing approaches primarily evolve C++ simulator code rather than reasoning directly about the microarchitectural designs it represents. As C++ code embodies hardware designs with several layers of abstraction, evolving it directly forces LLMs to reason through implementation details rather than the actual architectural mechanisms, hardware state, and decision rules that define a microarchitectural design.

To this end, we propose \ourtool, an LLM-guided methodology that generates new microarchitectural policies by extracting, integrating and adapting architectural concepts from published research. \ourtool{} extracts the underlying concepts from prior works into a structured knowledge library, which then guides the evolution of joint natural-language architectural specifications, rather than C++ simulator code. This approach addresses three key challenges: architectural representation in natural language, validation of candidate designs under hardware constraints, and the coordination of microarchitecture components using diagnostic feedback through co-evolution. 

We evaluate \ourtool{} across single-core and multicore configurations using SPEC CPU2017/2026, Google, GAP, and XSBench workloads, and compare specification-level evolution against code-level evolution across different search strategies. Under Claude Opus 5.5, \ourtool{} discovers designs achieving up to 4.8\% higher held-out single-core performance than the Pythia–Mockingjay baseline. A selected EvoX design achieves improvements of 3.4\% and 1.5\% under full and restricted memory bandwidth, respectively, although its multicore performance does not differ significantly from the baseline. The EvoX search evaluates 40 candidates over 2.2 days, with approximately \$320 in LLM usage at API list prices.

\end{abstract}

\maketitle 

\section{Introduction} 
\label{sec:introduction}

Microarchitectural components have been extensively studied for decades, yet developing new mechanisms remains challenging. Each new optimization requires extensive human expertise in microarchitecture, along with an intensive cycle of architectural reasoning, manual implementation, and simulation-based evaluation. Recent advances in large language models (LLMs) and agentic systems have created opportunities to automate microarchitectural design exploration~\cite{agenticarchitect,gupta2026archagent,wang2026agentdse,xiong2026microevo}. Unlike traditional approaches that primarily search predefined architectural configurations~\cite{bai2023archexplorer}, LLMs enable the capacity to generate, evaluate, and refine microarchitectural mechanisms while leveraging simulation feedback~\cite{agenticarchitect,gupta2026archagent,gonzalez2026archagentv2}.

For example, ArchAgent~\cite{gupta2026archagent} proposes an evolutionary coding agent that rewrites the C++ source of a last-level cache (LLC) replacement policy within an architectural simulator, starting from a source policy (Mockingjay)~\cite{shah2022mockingjay}. ArchAgent v2~\cite{gonzalez2026archagentv2} extended this approach to hardware prefetching at three cache levels, starting from the Berti~\cite{berti} and Pythia~\cite{bera2021pythia} prefetchers. Both report designs that outperform their starting points~\cite{gupta2026archagent,gonzalez2026archagentv2}. However, these winning candidates remain largely similar to the source policy, demonstrating refinement rather than introducing fundamentally new ideas. This is demonstrated within the four LLC policies that ArchAgent reports. Three policies keep Mockingjay's core algorithm while adding heuristics, tuning parameters, or changing the signature of its predictor, while the fourth remains similar to SHiP~\cite{wu2011ship}.
Moreover, ArchAgent v2 states that its changes provide ``limited novelty''~\cite{gonzalez2026archagentv2}. Although these advances demonstrate the potential of LLM-guided microarchitectural design exploration, existing LLM-guided evolutionary approaches still face challenges to effectively elicit conceptually unique microarchitectural candidates.

\begin{tcolorbox}[
    colback=rqbanner,
    colframe=black,
    boxrule=0.5pt,
    arc=5pt,
    left=6pt,
    right=6pt,
    top=3pt,
    bottom=3pt,
    before skip=3pt,
    after skip=3pt
]
\textbf{Research Question}

\smallskip
\textit{Can LLM-guided evolution through natural-language specifications discover novel, hardware-feasible microarchitectural policies beyond existing research?}
\end{tcolorbox}
\noindent\textbf{Challenge 1 (C1): Architectural representation and interpretability.} 
Existing code-level evolution approaches~\cite{gupta2026archagent,gonzalez2026archagentv2} directly modify C++ simulator implementations, where architectural mechanisms are intertwined with simulator-specific abstractions and implementation details. Consequently, it can be difficult to isolate the underlying hardware structures and decision rules that an evolved implementation actually models, complicating the assessment and validation of the proposed microarchitectural design. In contrast, structured natural-language specifications explicitly describe these architectural decisions, providing a reference for validation while potentially enabling LLMs to reason more effectively about high-level mechanisms during design exploration~\cite{wang2025planning}.
\textbf{Challenge 2 (C2): Hardware-constrained candidate validation.} Generated candidates must implement their intended architectural behavior and satisfy a hardware storage budget, but successful compilation and simulation do not establish either requirement. Simulator code is several layers of abstraction away from the hardware being designed and is not directly bound by its structure, sizes, or constraints. Therefore, the sizing of the underlying structures can be incorrectly modeled at the simulation level. Thus, validation requires an explicit representation of the intended architectural behavior, programmatic hardware storage accounting, and consistency checks between the architectural specification and the generated C++ simulator implementation.
\textbf{Challenge 3 (C3): Co-evolution of interacting microarchitectural components.} Interacting microarchitectural policies influence each other's performance, so a change that benefits one policy can negate the gain of another~\cite{gonzalez2026archagentv2}. Existing LLM-guided approaches evolve only one type of component at a time~\cite{gupta2026archagent,gonzalez2026archagentv2,agenticarchitect}, which, to our understanding, leaves the co-evolution of different components unexplored.

To address these challenges, we propose \ourtool{}, an LLM-guided methodology that evolves microarchitectural policies in natural language, from architectural ideas extracted from published research. Inspired by the process through which hardware architects study prior designs and formulate new mechanisms before C++ simulator code implementation, \ourtool{} constructs a structured idea library from published research papers, capturing their architectural mechanisms, observable signatures, hardware state, and decision rules. Our methodology addresses the three challenges through corresponding solutions. \textbf{Solution 1 (S1): Specification-level evolution (C1).} An LLM uses research-derived architectural knowledge and evaluation feedback to generate and refine natural-language architectural specifications, rather than directly evolving C++ simulator implementations. \textbf{Solution 2 (S2): Staged candidate validation (C2).} Each candidate undergoes architectural feasibility and hardware storage checks before C++ generation, followed by implementation-consistency checks before full simulation. \textbf{Solution 3 (S3): Coordinated evolution with diagnostic feedback (C3).} A joint specification represents interacting policies and their permitted information exchange, enabling simulation diagnostics to guide coordinated architectural revisions.

We demonstrate \ourtool through a case study that jointly evolves LLC replacement and L2 prefetching policies, including their permitted information exchange. These policies play important roles in memory-system performance, as the LLC replacement policy determines which cache lines are retained, while the L2 prefetching policy determines which cache lines to fetch in advance to hide memory latency. Their decisions are interdependent, as prefetching changes the cache lines available for replacement, while replacement determines whether prefetched lines remain in the cache long enough to be useful~\cite{demandmin,kim2017kpc}. This interdependence makes their co-design challenging and provides a meaningful case study for evaluating whether \ourtool can evolve coordinated architectural mechanisms from research-derived knowledge through natural-language specifications to improve overall performance.

The main contributions of this paper are:
\begin{itemize}
    \item \textbf{Specification-level evolution for microarchitectural design exploration.} 
    We introduce \ourtool, an LLM-guided methodology that extracts architectural concepts from published research to generate and evolve new microarchitectural policies through natural-language specifications rather than C++ simulator code.

    \item \textbf{Hardware-constrained architectural validation.} 
We introduce staged validation that assesses architectural feasibility, enforces hardware storage constraints, and checks specification–simulation consistency before evaluation.
    \item \textbf{Coordinated architectural evolution with diagnostic feedback.} 
    We jointly evolve LLC replacement and L2 prefetching policies through a shared architectural specification, using simulation diagnostics to guide revisions to their mechanisms and permitted interactions.
    \item \textbf{Extensive evaluation of evolution strategies.} 
     We compare against Pythia--Mockingjay~\cite{bera2021pythia,shah2022mockingjay}, the DPC-4~\cite{dpc4_2026} Pythia--DRRIP~\cite{jaleel2010rrip} baseline, and code-level evolution across SPEC CPU 2017~\cite{spec2017}, SPEC CPU 2026~\cite{madhav2026spec}, Google~\cite{google2022workloadtraces}, GAP~\cite{beamer2017gap}, and XSBench~\cite{tramm2014xsbench} workloads under matched search conditions, along with ablation studies across validation, microarchitectural policies, evolution strategies, and LLMs.

    \item \textbf{Performance and search efficiency.} Within 40 search candidates, \ourtool achieves 3.4\% higher geometric-mean (GM) IPC than Pythia--Mockingjay~\cite{bera2021pythia,shah2022mockingjay} and 1.9\% higher IPC than code-level evolution under matched conditions, on held-out traces at full memory bandwidth.

\end{itemize}

\section{Background}
\label{sec:background}

This section below highlights prior work in automated search for microarchitectural designs (Section~\ref{sec:background-evolution}), allowing with foundational baselines regarding LLC replacement and L2 prefetching upon which \ourtool is applied. (Section~\ref{sec:background-cache-policies}).

\subsection{Automated $\mu$-architectural Design Search}
\label{sec:background-evolution}

Automated search explores candidate designs by generating, evaluating, and selecting alternatives within a design space using approaches such as random search, hill climbing, and evolutionary search~\cite{eiben2015evolutionary,jimenez2017multiperspective}. Evolutionary search iteratively improves candidates through mutation (modifying an existing candidate) and recombination (combining information from multiple candidates)~\cite{eiben2015evolutionary}. A candidate representation defines which design decisions can be modified, such as architectural parameters, C++ simulator implementations, or architectural specifications, while a fitness function assigns performance scores, and design constraints determine candidate admissibility~\cite{eiben2015evolutionary}. Prior microarchitectural research has used genetic algorithms to explore insertion and promotion vectors for LLC replacement policies~\cite{jimenez2013ipv}, and Multiperspective Reuse Prediction combines randomized exploration with hill climbing to select predictor features~\cite{jimenez2017multiperspective}. Although these approaches demonstrate the effectiveness of automated search, their predefined candidate representations limit exploration to the architectural decisions they encode, thereby restricting the discovery of microarchitectural policies that could improve overall system performance.

LLM-guided evolutionary search uses LLMs to generate and refine candidate designs based on architectural requirements and evaluation feedback, enabling exploration of architectural mechanisms, decision rules, and hardware state beyond predefined design choices~\cite{gupta2026archagent,novikov2025alphaevolve,gonzalez2026archagentv2}. For example, ArchAgent~\cite{gupta2026archagent} and ArchAgent v2~\cite{gonzalez2026archagentv2} demonstrate LLM-guided evolutionary search for LLC replacement and hardware prefetching policies, respectively. Depending on the candidate representation, evolutionary search can modify architectural parameters, C++ simulator implementations, or architectural specifications. 

In contrast to code-level evolution which directly modifies C++ implementations, we propose specification-level evolution modifies the architectural specifications from which these C++ implementations are generated for evaluation. Furthermore, existing code-level approaches evolve one type of policy at a time, potentially limiting coordinated policy evolution, and intertwine architectural decisions with implementation details, as discussed further in Section~\ref{sec:motivation}. Consequently, realizing the broader potential of LLM-guided evolutionary search requires a candidate representation that states the hardware structures and decision rules of a design explicitly, rather than leaving them implicit in simulator code.

\begin{figure}[t]
    \centering
    \includegraphics[width=\linewidth]{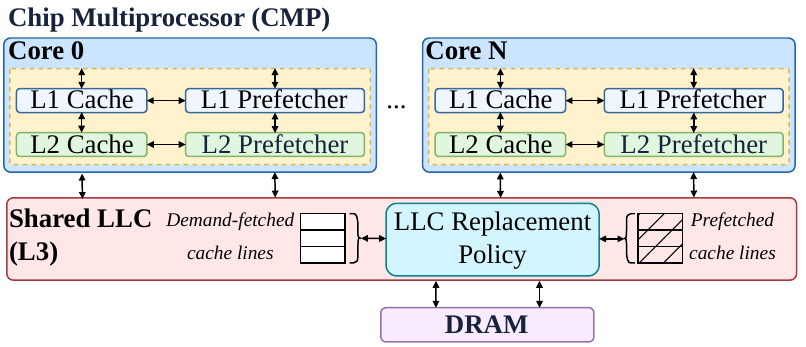}
    \caption{Interaction b/w LLC replacement
    and L2 prefetcher.}
    \label{fig:interaction}
\end{figure}

\subsection{LLC Replacement and L2 Prefetching}
\label{sec:background-cache-policies}

\noindent\textbf{LLC Replacement.} The shared LLC serves requests that miss in the upper cache levels, reducing accesses to main memory~\cite{jimenez2013ipv}. As illustrated in Figure~\ref{fig:interaction}, the LLC replacement policy determines which cache blocks are retained when capacity is limited. The LLC replacement policy makes three decisions: (i) the priority assigned to a cache block when it is inserted (insertion); (ii) how that priority changes upon a hit (promotion); and (iii) which block to evict when a cache set is full (victim selection)~\cite{jimenez2013ipv,jaleel2010rrip}. The policy maintains hardware state, such as recency information or re-reference predictions~\cite{jaleel2010rrip}, and may use predictors trained on past access behavior~\cite{jain2016hawkeye} to identify blocks likely to be reused. These decisions determine which cache blocks remain available for future accesses and directly influence LLC miss behavior.

\noindent\textbf{L2 prefetching.} An L2 prefetcher observes memory accesses reaching the L2 cache, predicts which cache blocks the program will request next, and fetches them in advance to hide memory latency~\cite{srinath2007fdp,bera2021pythia}. As shown in Figure~\ref{fig:interaction}, each core has an L2 prefetcher whose requests can bring prefetched cache lines into the shared LLC. The prefetching policy determines which blocks to fetch, how far ahead to fetch them (distance), and how many blocks to fetch per trigger (degree)~\cite{srinath2007fdp}. Three metrics characterize its effectiveness: (i) coverage, the fraction of demand misses eliminated by prefetching~\cite{bingo}; (ii) accuracy, the fraction of prefetched blocks accessed by demand requests before eviction~\cite{srinath2007fdp}; and (iii) timeliness, whether useful prefetches arrive before the corresponding demand accesses~\cite{berti}. These decisions affect not only how much memory latency is hidden but also the cache capacity and memory bandwidth consumed by prefetch requests.

\noindent\textbf{Coupled prefetching and replacement.} LLC replacement and L2 prefetching policies are interdependent because their decisions jointly influence cache contents and memory-system performance. L2 prefetching policy affects replacement decisions as prefetched cache blocks occupy LLC capacity and may displace blocks needed by subsequent demand accesses, causing cache pollution~\cite{srinath2007fdp,bera2021pythia}. Conversely, LLC replacement policy affects prefetching effectiveness by determining whether prefetched blocks remain in the LLC long enough to be used~\cite{demandmin,kim2017kpc}. Figure~\ref{fig:interaction} depicts how these policies influence each other's effectiveness in a multicore processor, where prefetch requests can also interfere with other applications by competing for shared LLC capacity and memory bandwidth~\cite{friendlyfire,ebrahimi_isca2011}. Consequently, improving either policy independently does not necessarily improve their combined performance, motivating the coordinated design of LLC replacement and L2 prefetching policies.

\noindent\textbf{Case-study rationale.} We select the co-design of LLC replacement and L2 prefetching policies as a demanding case study for evaluating \ourtool. Decades of research have produced sophisticated policies for both, including RRIP~\cite{jaleel2010rrip}, Hawkeye~\cite{jain2016hawkeye}, Mockingjay~\cite{shah2022mockingjay}, and Pythia~\cite{bera2021pythia}, as well as coordinated approaches such as PACMan~\cite{pacman} and KPC~\cite{kim2017kpc}. These established designs provide strong baselines that make further performance improvements challenging. Each policy must also satisfy a hardware storage budget, which restricts the set of feasible policy pairs. Existing LLM-guided approaches evolve one type of policy at a time~\cite{gupta2026archagent,agenticarchitect,gonzalez2026archagentv2}, and none demonstrates the joint evolution of LLC replacement and L2 prefetching policies. Finally, the SPEC CPU2017~\cite{spec2017}, SPEC CPU2026~\cite{madhav2026spec}, GAP~\cite{beamer2017gap}, XSBench~\cite{tramm2014xsbench}, and Google~\cite{google2022workloadtraces} trace suites are large enough for evaluation on held-out traces, which reveals whether a search overfits the workloads used during evolution~\cite{agenticarchitect}.

\begin{figure*}[t]
    \centering
    \includegraphics[width=\textwidth]{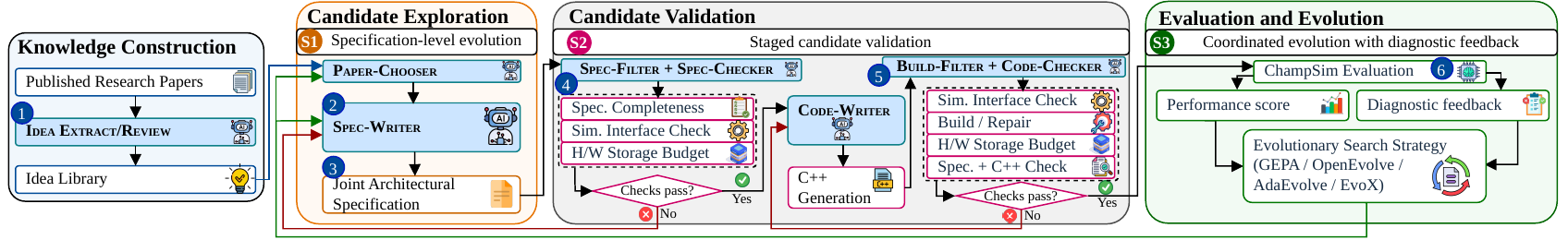}
    \caption{Overview of the \ourtool methodology.}
    \label{fig:framework}
\end{figure*}

\section{Motivation}
\label{sec:motivation}

\begin{table}[t]
    \centering
    \caption{Prior approaches vs. our \ourtool{} methodology.}
    \label{tab:prior-work}

    \newcommand{\priorYes}{\textcolor[HTML]{225EA0}{\cmark}}
    \newcommand{\priorNo}{\textcolor{red}{\xmark}}
    \newcommand{\priorUnknown}{\textcolor[HTML]{444444}{\ensuremath{\Delta}}}

    \scriptsize
    \setlength{\tabcolsep}{1.5pt}
    \renewcommand{\arraystretch}{1.2}

    \begin{tabularx}{\columnwidth}{
        @{}>{\raggedright\arraybackslash}p{0.29\columnwidth}
        *{5}{>{\centering\arraybackslash}X}@{}
    }
        \toprule
        Work
        & \makecell{Arch.\\Spec}
        & \makecell{Pre-sim.\\Check}
        & \makecell{H/W\\Storage\\Check}
        & \makecell{Diagnostic\\F/B}
        & \makecell{Joint-\\Policy\\Evolution} \\
        \midrule

        IPV~\cite{jimenez2013ipv}
        & \priorNo & \priorUnknown & \priorUnknown & \priorUnknown & \priorNo \\

        Multiperspective~\cite{jimenez2017multiperspective}
        & \priorNo & \priorUnknown & \priorUnknown & \priorUnknown & \priorNo \\

        PACIPV~\cite{mostofi2025pacipv}
        & \priorNo & \priorUnknown & \priorUnknown & \priorUnknown & \priorNo \\

        \midrule

        AgentDSE~\cite{wang2026agentdse}
        & \priorUnknown & \priorUnknown & \priorUnknown & \priorUnknown & \priorNo \\

        MicroEvo~\cite{xiong2026microevo}
        & \priorUnknown & \priorUnknown & \priorUnknown & \priorUnknown & \priorNo \\

        ArchAgent~\cite{gupta2026archagent}
        & \priorNo & \priorYes & \priorUnknown & \priorUnknown & \priorNo \\

        ArchAgent v2~\cite{gonzalez2026archagentv2}
        & \priorNo & \priorYes & \priorYes & \priorUnknown & \priorNo \\

        Agentic Architect~\cite{agenticarchitect}
        & \priorNo & \priorYes & \priorYes & \priorUnknown & \priorNo \\

        Presage~\cite{giordano2026presage}
        & \priorNo & \priorUnknown & \priorUnknown & \priorUnknown & \priorNo \\

        \midrule

        \textbf{\ourtool{} (our work)}
        & \priorYes & \priorYes & \priorYes & \priorYes & \priorYes \\

        \bottomrule
    \end{tabularx}

    \vspace{3pt}

    \begin{minipage}{\columnwidth}
        \scriptsize
        \priorYes~Supported \quad
        \priorNo~Outside the described approach \quad
        \priorUnknown~Unspecified
    \end{minipage}
\end{table}

Developing new microarchitectural policies often involves studying prior research, identifying useful architectural ideas, and adapting them to construct new designs~\cite{bera2021pythia,shah2022mockingjay,berti}. Architects reason about architectural mechanisms, hardware state, and decision rules before translating their designs into C++ implementations. Inspired by this process, we use LLMs to extract and integrate architectural ideas from published research to generate microarchitectural policies, rather than directly optimizing existing C++ implementations.

Generating architectural ideas across different designs requires understanding their underlying mechanisms and dependencies. Existing policies use indexing features, hardware state, and decision rules to govern their behavior~\cite{bera2021pythia,shah2022mockingjay,berti}. For example, adapting a prediction mechanism from one policy may require changes to the hardware state used to maintain its history or to the decision rules that consume its predictions~\cite{bera2021pythia,shah2022mockingjay}. Therefore, combining mechanisms involves more than selecting individual ideas; their inputs, state updates, and behavior must remain architecturally consistent. For interacting policies, their permitted information exchange must also be explicitly represented to support coordinated architectural decisions~\cite{pacman,kim2017kpc}. These requirements motivate a joint natural-language architectural specification that captures individual policy behavior and cross-policy interactions independently of simulator-specific implementation details.

Table~\ref{tab:prior-work} compares existing microarchitectural design approaches across architectural specification, pre-simulation validation, hardware storage accounting, diagnostic feedback, and joint policy evolution. Traditional automated approaches, including insertion and promotion vector search~\cite{jimenez2013ipv} and Multiperspective Reuse Prediction~\cite{jimenez2017multiperspective}, explore predefined architectural choices. Such representations enable systematic exploration of selected mechanisms but constrain the architectural decisions that can be modified. More recent LLM-guided approaches, including ArchAgent~\cite{gupta2026archagent} and ArchAgent v2~\cite{gonzalez2026archagentv2}, can generate and refine simulator code beyond predefined parameter spaces. However, as summarized in Table~\ref{tab:prior-work}, these approaches evolve simulator code rather than explicitly maintaining and evolving architectural specifications that separate policy mechanisms from C++ code. Consequently, the representation used for evolution affects not only how candidates are modified but also how architectural ideas can be examined and combined.

A second consideration is if generated designs satisfy hardware constraints. Combining mechanisms from different policies may introduce unbuildable designs due to unrestrained hardware incompatible assumptions, including additional hardware state, incompatible assumptions about the available features, or inconsistencies between intended behavior and generated C++ simulator implementations. These issues are particularly important when the resulting design must satisfy a fixed hardware storage budget~\cite{bera2021pythia,shah2022mockingjay}. Table~\ref{tab:prior-work} distinguishes pre-simulation validation from hardware storage accounting because the two address different requirements. For example, ArchAgent v2~\cite{gonzalez2026archagentv2} employs pre-simulation checks and hardware storage estimation, but these capabilities alone do not establish that generated C++ implementations preserve the intended architectural behavior. An explicit architectural specification provides a reference for checking this correspondence. Thus, architectural idea generation requires not only the ability to generate new mechanisms but also a means to validate their feasibility and their consistency with the specification before incurring the cost of expensive simulation.

Generating new policies becomes more challenging when multiple microarchitectural policies influence one another, as LLC replacement and L2 prefetching do (Section~\ref{sec:background-cache-policies}). Table~\ref{tab:prior-work} highlights joint policy evolution and diagnostic feedback as distinct capabilities. Although prior approaches demonstrate automated policy optimization, including prefetching across multiple cache levels~\cite{gupta2026archagent,gonzalez2026archagentv2}, jointly generating different policy types requires considering both their individual mechanisms and their interactions. Moreover, aggregate performance scores indicate whether a policy pair improves performance but provide limited information about which architectural decisions account for the observed behavior. Diagnostic feedback can help identify these effects and guide subsequent refinement of the joint design~\cite{srinath2007fdp,bai2023archexplorer}.

Prior approaches explore predefined architectural choices or evolve C++ simulator code~\cite{jimenez2013ipv,jimenez2017multiperspective,gupta2026archagent,gonzalez2026archagentv2}, but do not explicitly generate new microarchitectural policies by extracting and adapting architectural ideas from published research within an evolving architectural specification. This distinction motivates a different design methodology, in which architectural ideas and their interactions, rather than C++ implementations, become the objects of evolution. Realizing this methodology requires a joint natural-language architectural specification that represents the generated mechanisms (\textbf{C1}), staged validation that preserves architectural intent and enforces hardware constraints (\textbf{C2}), and diagnostic-guided refinement of interacting policies (\textbf{C3}). Accordingly, \ourtool{} focuses on generating new microarchitectural designs from research-derived ideas, with code generation and simulation serving to validate and evaluate those designs rather than guide the evolution process.

\newcommand{\chooser}{\textsc{Paper-Chooser}}
\newcommand{\specwriter}{\textsc{Spec-Writer}}
\newcommand{\specgate}{\textsc{Spec-Filter}}
\newcommand{\checkf}{\textsc{Spec-Checker-F}}
\newcommand{\checkd}{\textsc{Spec-Checker-D}}
\newcommand{\checks}{\textsc{Spec-Checker-S}}
\newcommand{\codewriter}{\textsc{Code-Writer}}
\newcommand{\buildgate}{\textsc{Build-Filter}}
\newcommand{\codechecker}{\textsc{Code-Checker}}
\newcommand{\ideaextract}{\textsc{Idea Extract/Review}}

\section{Proposed Methodology}
\label{sec:proposed-methodology}

Given the challenges identified in Section~\ref{sec:motivation}, we introduce \ourtool{}, a methodology for co-evolving microarchitectural policies from architectural ideas extracted from published research papers. \ourtool{} decomposes published designs into reusable architectural mechanisms, indexing features, hardware state, and decision rules, then adapts and combines these ideas to generate new policies. Unlike code-level evolution, \ourtool{} represents and evolves these generated policies through joint natural-language architectural specifications rather than directly modifying C++ simulator implementations. This section describes the research-derived idea library, the architectural representation, the generation process, and the validation mechanisms underlying this methodology (Figure~\ref{fig:framework}).

\subsection{Overview}

As illustrated in Figure~\ref{fig:framework}, \ourtool{} organizes its workflow into knowledge construction, candidate exploration (S1), candidate validation (S2), and evaluation and evolution (S3). In the knowledge construction stage, \ourtool extracts and organizes prior architectural ideas into a structured library ($\mathcal{L}$), retaining relationships between mechanisms and their architectural elements. During candidate exploration, the \chooser{} retrieves relevant research-derived architectural knowledge from the idea library, and the \specwriter{} adapts and combines these ideas to generate new policies represented by a joint natural-language architectural specification, $D=(D_R,D_P,D_X)$, where $D_R$ and $D_P$ describe the LLC replacement and L2 prefetching policies, respectively, and $D_X$ specifies their permitted coordination. Each policy explicitly describes its architectural idea, hardware state, and decision rules, allowing \ourtool{} to refine individual mechanisms and their interactions independently of simulator-specific implementation details.

\ourtool{} validates each generated architectural specification before translating it into a C++ simulator implementation. The \specgate{} and Spec-Checker assess architectural feasibility and hardware storage budget compliance, after which the \codewriter{} generates a C++ implementation that is checked by the \buildgate{} and \codechecker{} for implementation constraints and consistency with the specification. The specification remains the authoritative design representation throughout this process. Validated C++ implementations are evaluated in ChampSim to obtain performance scores and architectural diagnostic feedback, which guide subsequent specification generation under an external evolutionary strategy. Thus, architectural specifications, rather than C++ code, remain the objects of exploration and evolution in \ourtool{}.

\noindent\textbf{Methodology Components.} \ourtool{} implements its LLM-based operations through task-specific \textit{agents}, each consisting of an LLM invocation with predefined instructions and a context template populated with relevant inputs. These include the \chooser{}, \specwriter{}, \checkf{}, \checkd{}, \checks{}, \codewriter{}, and \codechecker{}. In contrast, \textit{filters}, such as \specgate{} and \buildgate{}, are Python validation scripts that enforce predefined rules without LLM involvement. Table~\ref{tab:caper-agents} summarizes the agents and filters used throughout the methodology.

\begin{table}[t]
\centering
\caption{\ourtool{} agents and validation filters.}
\label{tab:caper-agents}
\footnotesize
\setlength{\tabcolsep}{3pt}
\renewcommand{\arraystretch}{1.08}
\begin{tabularx}{\columnwidth}{@{}llX@{}}
\toprule
\textbf{Component} & \textbf{Type} & \textbf{Responsibility} \\
\midrule
\ideaextract{} & LLM & Extracts and reviews ideas from published papers \\
\textsc{Paper-Chooser} & LLM & Selects relevant research papers \\
\textsc{Spec-Writer} & LLM & Generates/revises architectural specifications \\
\midrule
\textsc{Spec-Filter} & Python & Checks specification format and storage budget \\
\textsc{Spec-Checker-F} & LLM & Checks architectural feasibility \\
\textsc{Spec-Checker-D} & LLM & Checks soundness, coordination, and novelty \\
\textsc{Spec-Checker-S} & LLM & Checks fidelity to source papers \\
\midrule
\textsc{Code-Writer} & LLM & Generates/repairs C++ implementations \\
\textsc{Build-Filter} & Python & Checks files, compilation, and hardware storage \\
\textsc{Code-Checker} & LLM & Checks spec. and C++ implementation consistency \\
\bottomrule
\end{tabularx}
\end{table}

\subsection{Research-Derived Architectural Knowledge}

\ourtool{} first constructs a structured library of architectural knowledge extracted from 177 microarchitectural research publications. As illustrated in Figure~\ref{fig:framework}\,\figcallout{1}, rather than treating each work as a single design concept, \ourtool{} decomposes each publication into its underlying mechanisms, indexing features, hardware structures, stored information, and decision rules. This library remains fixed and enables architectural concepts from different designs to be later retrieved and adapted during subsequent policy generation.

For each paper, the \ideaextract{} agent uses an LLM to extract a design record that describes its core mechanism, rationale, indexing features, hardware state and storage, state updates, decision rules, parameters, reported results, and limitations. The same agent uses another LLM to review each record against its source paper, retaining or revising the extracted information. In the current corpus, 101 records were retained and 76 revised. This review improves alignment with the source material.

To identify related architectural concepts across designs, \ourtool{} organizes the extracted records into three linked catalogs, $\mathcal K=(\mathcal S,\mathcal T,\mathcal C)$, where \(\mathcal S\) contains canonical indexing features, \(\mathcal T\) describes hardware state structures, and \(\mathcal C\) contains canonical stored-content names. Each structure records its indexing method or cache location, stored fields, and storage dimensions. Catalog entries preserve their originating designs and the architectural purpose of each element, allowing mechanisms to be retrieved through shared features and state characteristics (Table~\ref{tab:idea-library}).

\begin{table}[t]
\centering
\caption{Excerpt from \ourtool{}'s idea library.}
\label{tab:idea-library}
\footnotesize
\setlength{\tabcolsep}{3pt}
\renewcommand{\arraystretch}{1.0}
\begin{tabularx}{\columnwidth}{@{}lX@{}}
\toprule
\textbf{Entry} & \textbf{Description} \\
\midrule
\multicolumn{2}{@{}l}{\textbf{Indexing features}} \\
instruction pc &
Observed by 90 designs; indexes history (Berti), informs cache-friendliness prediction (Hawkeye), and forms predictor signatures (Mockingjay). \\
\midrule
\multicolumn{2}{@{}l}{\textbf{Structures}} \\
Berti~\cite{berti} &
PC tag, search count, stride, coverage counter, delta status, FIFO pointer. \\
Hawkeye~\cite{jain2016hawkeye} &
Confidence counter. \\
Mockingjay~\cite{shah2022mockingjay} &
Reuse interval; also indexed by hit/miss, request type, and core identity. \\
\midrule
\multicolumn{2}{@{}l}{\textbf{Stored content}} \\
confidence counter &
Confidence in a learned prediction. \\
coverage counter &
Frequency of candidate-delta coverage. \\
reuse interval &
Reuse separation in access-count or time units. \\
\bottomrule
\end{tabularx}
\end{table}

\ourtool{} constructs these catalogs incrementally, reusing canonical names when their meanings match and introducing new names otherwise. Information not explicitly available in a source record is marked as inferred rather than attributed to the original design. Programmatic checks enforce identifier coverage, uniqueness, and valid references across the constructed catalogs. We note that these checks establish structural consistency; however, semantic fidelity to the source papers remains dependent on the LLM-based extraction and independent source-comparison review.

During candidate exploration, the idea library $L$ supplies the architectural knowledge used in the generation context $c_t$. The \chooser{} consults the catalogs in $K$ to select research papers relevant to the planned architectural revision, and the \specwriter{} receives their full design records together with the catalogs. This evidence supports subsequent assessments of architectural similarity and novelty. Thus, knowledge construction provides a consistent, research-derived context for proposing and reviewing architectural ideas without restricting evolution to policy implementations.

\subsection{Joint Architectural Representation}

To address C1, \ourtool{} represents candidate policies through structured natural-language architectural specifications, separating architectural intent from C++ simulator implementation details. As depicted in Figure~\ref{fig:framework}\,\figcallout{3}, each candidate explicitly describes the underlying architectural mechanisms, hardware state, and decision rules of the policies being designed, rather than evolving C++ code.

We denote each policy \(D_i\) in \(i\in\{R,P\}\) as
$D_i=(I_i,H_i,B_i)$,
where \(I_i\) describes the architectural idea, \(H_i\) specifies the hardware state, and \(B_i\) defines the policy behavior. These components form three structured parts of the natural-language specification, providing a consistent representation of the architectural policies being evolved. The \textit{idea} (\(I_i\)) describes the observation the policy exploits and its closest prior designs, with what it takes from each and how it differs. The \textit{state} (\(H_i\)) declares the hardware structures required by the mechanism, including their stored contents, organization, indexing, capacities, field widths, and initialization, and ends with a storage table with one row per structure. The \textit{behavior} (\(B_i\)) specifies what the policy does at each moment the simulator reports, its learning and decision rules, and all parameters, using pseudocode when precise behavior is required. For LLC replacement, these rules govern insertion, promotion, bypass, and victim selection. On the other hand, for L2 prefetching, they govern access tracking, address generation, and prefetch request control.

Beyond the individual policies, the cooperation component \(D_X\) specifies how information conveyed through prefetcher metadata influences LLC replacement decisions, or explicitly declares the policies independent. The available communication interface is fixed by defined rules in the agent context, while evolution determines how the two policies use it.

This representation enables research-derived architectural ideas to be integrated into complete policy mechanisms, while preserving their rationale, state requirements, and interactions (C1). The same specification serves as the reference for hardware-constraint enforcement and implementation-consistency checks (C2).

\subsection{Research-Guided Architectural Generation}\label{sec:research-guided}

\ourtool{} generates architectural policies by adapting ideas extracted from prior research into joint policy specifications. An initial seed \(D_0\) is constructed from mechanisms retrieved from the idea library \(\mathcal L\). Subsequent candidates build on previously evaluated specifications selected by an evolutionary strategy \(\sigma\), without inheriting their C++ implementations. This separates the generation of architectural mechanisms from the underlying search strategy.

\noindent\textbf{Research-Guided Retrieval.} At iteration $t$, let $p_t$ denote the parent specification(s) selected by the external evolutionary strategy $\sigma$, and $\mathcal{H}_t$ the relevant evaluation history and search guidance. The \chooser{} retrieves a subset of research records from the idea library $L$ based on these inputs, $R_t = \operatorname{Retrieve}(L,p_t,\mathcal{H}_t) \subseteq L$.

The \chooser{} consults the architectural catalogs $K$ to identify relevant mechanisms and closely related published designs. It considers both policy domains and favors less frequently consulted sources to diversify the architectural concepts available for generation and novelty assessment.

\noindent\textbf{Architectural Generation.} \ourtool{} constructs the generation context $c_t$ from the parent specification(s) selected by the evolutionary strategy, research records $R_t$ retrieved by the \chooser{}, architectural catalogs $K$, prior designs, evaluation feedback, and hardware constraints. As depicted in Figure~\ref{fig:framework}\,\figcallout{2}, the \specwriter{} uses this context to generate a revised architectural specification, $D_{t+1}\sim G_\theta(\cdot\mid c_t)$, where $G_\theta$ denotes the LLM generator.

The generated architectural specification may revise the architectural idea \(I_i\), hardware state \(H_i\), and behavior \(B_i\) of either policy, together with their permitted coordination \(D_X\). Thus, architectural specification generation can modify both individual mechanisms and their interactions within a single candidate.

\noindent\textbf{Novelty-Aware Generation.} \ourtool{} distinguishes reusable architectural concepts from complete policy mechanisms. Individual elements from prior research, such as features, indexing methods, and stored fields, may be adapted to construct new mechanisms. However, newly proposed core mechanisms must not simply reproduce published designs, modify their parameters, or directly combine existing policies without a distinct architectural mechanism. Previously evolved designs are also considered when assessing novelty, although candidates may retain and refine their parents' mechanisms, except when deriving directly from the literature-based seed. These criteria guide specification generation and are assessed during the subsequent validation stage.

This process enables research-derived architectural ideas to be generated and refined at the specification level, while leaving the candidate evolution process to the external search strategy.

\subsection{Hardware-Constrained Translation and Validation}

To address C2, \ourtool{} employs two stages of candidate validation, as shown in Figure~\ref{fig:framework}. Before C++ generation, the \specgate{} and \checkf{}/D/S validate the architectural specification for completeness, feasibility, soundness, and compliance with the hardware storage budget. After the \codewriter{} generates a C++ implementation, the \buildgate{} and \codechecker{} check its compilation, simulator compatibility, hardware budget compliance, and consistency with the specification before full simulation.

\noindent\textbf{Specification Validation.} As showcased in Figure~\ref{fig:framework}\,\figcallout{4}, the \specgate{} first checks the completeness and structural consistency of each specification $D$, including the declared hardware state $H_i$ for each policy $i\in\{R,P\}$, and verifies compliance with the hardware storage budget. The required storage is calculated from the specification as $M_i(D)=\sum_{h\in H_i}n_hw_h\leq M_i^{\max}$, where $n_h$ denotes the number of instances of structure $h$, including replication, $w_h$ is its declared width in bits, and $M_i^{\max}$ is the policy's hardware storage budget. All storage quantities in this constraint are expressed in bits, with the reported KB budgets converted accordingly. The LLM-based \checkf{} and \checkd{} then assess architectural feasibility, including available features, soundness of state updates and decision rules, permitted inter-policy interactions, consistency with referenced research, and novelty criteria established in Section~\ref{sec:research-guided}. The literature-derived seed is instead reviewed by \checks{} for fidelity to its source mechanisms. Specifications that fail these checks are revised by the \specwriter{} using reviewer feedback, at most twice, or rejected.

\begin{table}[t]
\centering
\caption{Hardware storage primitives.}
\label{tab:storage-primitives}
\footnotesize
\setlength{\tabcolsep}{2.5pt} 
\renewcommand{\arraystretch}{1.08}
\begin{tabularx}{\columnwidth}{@{}l l X@{}}
\toprule
\textbf{Primitive} & \textbf{Description} & \textbf{Example} \\
\midrule
\texttt{uint<W>} & Unsigned $W$-bit field & 52-bit page tag \\
\texttt{sint<W>} & Signed $W$-bit field & 5-bit price offset \\
\texttt{bits<W>} & Bit vector ($W>64$) & 320-bit vote vector \\
\texttt{table<E,N,K>} & $N$ entries, $K$ ways & 384-entry, 6-way page table \\
\texttt{per\_block<E>} & Per-cache-block state & 3-bit phase countdown per LLC block \\
\bottomrule
\end{tabularx}
\end{table}

\noindent\textbf{Translation to C++.} A specification passing specification-level validation is translated into a C++ simulator implementation $C_D$ by the \codewriter{}. The \codewriter{} must preserve the specified behavior $B_i$, hardware state $H_i$, and coordination rules $D_X$, while conforming to the simulator's permitted interfaces and execution events. Hardware storage budget compliance is enforced by construction, because the build process generates each module's persistent state directly from its specification tables using the fixed-width primitives of Table~\ref{tab:storage-primitives}. Compile-time checks verify the declared storage size, while source- and object-level checks reject additional or disallowed persistent state. As shown in Figure~\ref{fig:framework}\,\figcallout{5}, the \buildgate{} checks file scope and code-length limits, compiles the C++ code, verifies the hardware storage budget constraints, and runs a single-trace simulation check. Failures are returned to the \codewriter{} for at most two repair attempts.

\noindent\textbf{Consistency Check.} The \codechecker{} compares the generated C++ code $C_D$ against its architectural specification $D$, identifying missing, additional, or altered behavior and incorrect placement of operations within simulator interfaces. Detected defects are returned to the \codewriter{} for one repair attempt. Candidates that still fail the subsequent consistency check are rejected. If the \codewriter{} identifies an architectural rule that cannot be implemented as specified, the \specwriter{} first revises $D$, which must then pass specification-level validation again before C++ generation resumes.

Let $V_{\mathrm{spec}}(D)$ and $V_{\mathrm{code}}(D,C_D)$ denote acceptance by the specification and code-level validation stages, respectively. The admissible design space is $\mathcal{D}_{\mathrm{adm}}=\{D\mid V_{\mathrm{spec}}(D)=1,\ V_{\mathrm{code}}(D,C_D)=1\}$. Together, the \specgate{}, \checkf{}/D/S, \buildgate{}, and \codechecker{} combine programmatic constraint enforcement with LLM-based architectural and implementation-consistency assessments. Only candidates passing both stages proceed to ChampSim evaluation, ensuring that evaluated C++ code implementations correspond to validated architectural specifications and satisfy the enforced hardware storage budget constraints.

\subsection{Performance and Diagnostic Feedback}

To address C3, \ourtool{} combines performance rewards with architectural diagnostic feedback to guide the evolution of interacting policies. As demonstrated in Figure~\ref{fig:framework}\,\figcallout{6}, each validated specification $D$ is evaluated in ChampSim by simulating its generated C++ implementation $C_D$ on the training workloads, producing a performance score and architectural diagnostic feedback.

\noindent\textbf{Performance Reward.} Let \(\mathcal W\) denote the training programs, \(\mathcal T_w\) the simulation program traces \(w\), and \(\alpha_{wj}\) their normalized SimPoint~\cite{sherwood2002simpoint} weights, where \(\sum_{j\in\mathcal T_w}\alpha_{wj}=1\). The weighted cycles per instruction (CPI) of each program and the performance reward are defined as:
$\bar c_w(D)=\sum_{j\in\mathcal T_w}\alpha_{wj}\,\mathrm{CPI}_{wj}(D),$
with the final reward defined as
$r(D)=\left(\prod_{w\in\mathcal W}
\frac{\bar c_w(D_{\mathrm{ref}})}{\bar c_w(D)}
\right)^{1/|\mathcal W|}$
, where \(D_{\mathrm{ref}}\) denotes the fixed reference design. The reward is the geometric mean of per-program speedups, assigning equal weight to each program regardless of its number of traces. Candidates that fail simulation on any training trace receive no score. The evolutionary strategy \(\sigma\) uses these rewards for candidate selection.

\noindent\textbf{Architectural Diagnostic Feedback.} Beyond aggregate performance, \ourtool{} collects diagnostic observations \(\mathbf z(D)\) characterizing the behavior of both policies and their effects on the memory system. These include LLC demand misses and eviction behavior, prefetch coverage, accuracy, timeliness, DRAM traffic, bandwidth utilization, CPI, and memory-stall breakdowns. \ourtool{} compares these measurements against the reference design, summarizing behavior across benchmark suites and retaining detailed observations for poorly performing traces and individual programs.

\noindent\textbf{Feedback-Guided Evolution.} The evolutionary search strategy uses performance rewards to select parent specifications for subsequent candidates, while diagnostic feedback is incorporated into the evaluation feedback $\mathcal{H}_t$. As showcased in Figure~\ref{fig:framework}, the \chooser{} uses this feedback to retrieve relevant architectural ideas, while the \specwriter{} uses the selected parent specifications, retrieved research, and diagnostic observations to propose targeted revisions to replacement ($D_R$), prefetching ($D_P$), or their information exchange ($D_X$). Thus, \ourtool{} uses diagnostic feedback to guide coordinated architectural specification revisions while retaining joint performance as the sole evolutionary objective.

\begin{table}[t]
    \centering
    \caption{Experimental configuration.}
    \label{tab:experimental-setup}
    \footnotesize

    \begin{tabularx}{\columnwidth}{
        >{\raggedright\arraybackslash}X
        >{\raggedright\arraybackslash}X
    }
        \toprule
        \textbf{Parameter} & \textbf{Configuration} \\
        \midrule

        \multicolumn{2}{l}{\textbf{Hardware and simulator}} \\
        Simulator and revision             & ChampSim~\cite{gober2022champsim}, DPC-4\cite{dpc4_2026} release, with CPI-stack~\cite{cpistack} counters added \\
        Core count and processor model      & 1 out-of-order core, or 4 such cores in the four-core configuration; 4\,GHz \\
        Fetch / issue / retire width        & 8 / 8 / 8 \\
        Reorder buffer size                 & 576 entries; load and store queues 240 and 112 \\
        L1 instruction cache                & 64\,KB, 8-way, 4 cycles \\
        L1 data cache                       & 48\,KB, 12-way, 5 cycles \\
        L2 cache                            & 2\,MB per core, 16-way, 10 cycles \\
        Shared LLC                          & 3\,MB (four-core 12\,MB), 12-way, 35 cycles \\
        Cache-line size                     & 64\,B \\
        MSHRs and queue capacities          & MSHRs 8 (L1I), 16 (L1D), 32 (L2), 64 (LLC; four-core 256); read, write and prefetch queues 32, 32, 16 (L2) and 32, 32, 32 (LLC; four-core 128 each); DRAM read and write queues 64 each \\
        Memory configuration and bandwidth  & 1 DDR channel, shared by the cores in the four-core configuration; 8\,B wide, 8 bank groups of 4 banks; 4800\,MT/s (38.4\,GB/s peak), or 800\,MT/s (6.4\,GB/s peak) in the restricted-bandwidth configuration; tCAS, tRCD, tRP 15\,ns \\
        Fixed policies at other levels      & Berti~\cite{berti} at L1D, no L1I prefetcher; LRU at L1I, L1D and L2; perceptron branch predictor~\cite{perceptron} \\
        L2 prefetcher storage budget        & 128\,KB per core\\
        LLC replacement storage budget      & 48\,KB for the 3\,MB LLC\\
     \bottomrule
    \end{tabularx}
\end{table}

\section{Evaluation}
\label{sec:evaluation}

\subsection{Experimental Setup}
\label{sec:experimental-setup}

\noindent\textbf{Simulation Environment.} We evaluate \ourtool{} using a modified ChampSim~\cite{gober2022champsim} with the single-core configuration of the 2026 Data Prefetching Championship (DPC-4)~\cite{dpc4_2026}, summarized in Table~\ref{tab:experimental-setup}. In our case study, \ourtool{} jointly evolves L2 prefetching and LLC replacement policies, using Pythia~\cite{bera2021pythia} and Mockingjay~\cite{shah2022mockingjay} as baselines. The L2 prefetcher and LLC replacement policy have hardware storage budgets of 128~KB and 48~KB, respectively.

\noindent\textbf{Workloads.} We use SPEC CPU2017~\cite{spec2017}, SPEC CPU2026~\cite{madhav2026spec}, GAP graph kernels~\cite{beamer2017gap}, XSBench~\cite{tramm2014xsbench}, and Google datacenter traces~\cite{google2022workloadtraces}. We retain SimPoint traces~\cite{sherwood2002simpoint} with at least 10\% program weight and more than one LLC miss per thousand instructions (MPKI), measured over 200M instructions following a 50M-instruction warm-up with prefetching disabled and LRU replacement. We split the traces within each program: the training set contains 105 traces from 75 programs, while the held-out set contains 102 traces from 68 programs. No trace appears in both sets, and 52 of the 68 held-out programs also have other traces in the training set.

\noindent\textbf{Hardware Configurations.} The single-core and multicore configurations are summarized in Table~\ref{tab:experimental-setup}. For sensitivity evaluation to memory bandwidth, we use a restricted-bandwidth configuration at 800~MT/s with unchanged DRAM latencies. For multicore evaluation, we use a four-core configuration with private 2~MB L2 caches and prefetchers, a shared 12~MB LLC, and one 4800~MT/s DRAM channel. In this configuration, we simulate 100 mixes of four held-out traces drawn at random, with 20M warm-up and 50M measured instructions per core.

\noindent\textbf{Evaluation Metrics.} We report IPC speedup relative to the Pythia-Mockingjay baseline, using SimPoint-weighted CPI for programs with multiple traces. Program-level speedups are aggregated using the GM, assigning equal weight to each program. For the four-core configuration, we report the harmonic speedup of each mix, namely the harmonic mean over the four cores of the IPC in the mix divided by the IPC of the same trace running alone with the DPC-4 baseline policies (Pythia + DRRIP). We give it relative to Pythia + Mockingjay and aggregate it over the mixes with the GM.

\subsection{Controlled Search Protocol}
\label{sec:search-protocol}

\noindent\textbf{Search Configuration.} We conduct evolutionary search using SkyDiscover~\cite{liu2026skydiscover} with GPT-6 Astra~\cite{openai2026gpt6astra} and Claude Opus 5.5~\cite{anthropic2026opus55}. We employ four evolutionary strategies: GEPA~\cite{agrawal2025gepa}, OpenEvolve~\cite{sharma2025openevolve}, AdaEvolve~\cite{cemri2026adaevolve}, and EvoX~\cite{liu2026evox}. We compare specification-level evolution against code-level evolution under matched evolutionary strategies, LLMs, workloads, and candidate budgets. Both approaches use the same idea library, applicable build checks, hardware storage constraints, diagnostic feedback, and performance objective, enabling a controlled comparison of candidate representations.

\noindent\textbf{Candidate Budget and Selection.} Each proposed candidate counts as one search attempt, including revisions and repairs. We allow up to two specification revisions, two repairs of the C++ simulator model following build or run failures, and one additional edit for a mismatch between the specification and the model. Candidates that remain unsuccessful receive no performance score but still count toward the search budget. The search objective is the GM of program-level IPC speedups across the 75 training programs on the single-core 4800~MT/s configuration. During a run, a candidate is marked when its training score exceeds that of the seed and of every earlier marked candidate by more than 0.001. We call the last marked candidate of a run its champion design, and we evaluate it on the held-out traces at 4800 and 800~MT/s.

\subsection{Results}
\label{sec:results}

We evaluate \ourtool{} to assess its ability to discover novel, hardware feasible, and high-performing designs through specification-level evolution. We examine the novelty and performance of the discovered designs, the impact of candidate representation on design quality, search efficiency, and computational cost, and the effectiveness of staged candidate validation. We also evaluate the contributions of the evolved LLC replacement and L2 prefetching policies, their interactions, and performance across workloads and hardware configurations in our case study. Below, we address the following questions:

\begin{itemize}
    \item \textbf{RQ1: Novelty.} How does the design discovered by \ourtool{} differ from its seed and published designs?
    \item \textbf{RQ2: Evolution Efficacy.} How effectively does \ourtool{} discover high-performing microarchitectural designs through specification-level evolution?
    \item \textbf{RQ3: Search Representation.} How does specification-level evolution compare with code-level evolution in design quality, search efficiency, and computational cost?
    \item \textbf{RQ4: Candidate Validation.} How effectively does \ourtool{}'s candidate validation enforce architectural feasibility and hardware storage constraints before simulation?
\end{itemize}

\subsubsection{\textbf{Case Study: Seed and Champion Design.}}

Starting from published designs can lead to incremental variations with limited novelty~\cite{gupta2026archagent,gonzalez2026archagentv2}. Instead, \ourtool{} constructs its seed design by combining architectural ideas retrieved from the idea library, rather than directly adopting an existing policy. The resulting seed integrates complementary prediction mechanisms across the cache hierarchy. Its L2 prefetcher predicts future accesses using agreement among recently observed page footprints, while its LLC replacement policy predicts block reuse based on the requesting instruction. The prefetcher places high-confidence predictions in the L2 and lower-confidence predictions in the LLC, using prefetch metadata to distinguish them for replacement decisions.

For detailed architectural analysis, we select the EvoX champion generated using Claude Opus 5.5 at candidate 38, following 17 mutation steps. Although AdaEvolve and GEPA achieve higher single-core held-out speedups, AdaEvolve shows no gain on four-core mixes, while GEPA's 3.6\% improvement is concentrated in Google traces (8.8\%). In contrast, EvoX's evolved L2 prefetcher retains a 2.3\% gain on four-core mixes with Mockingjay, motivating its selection for examining architectural novelty and cross-configuration behavior. We compare all search strategies in RQ3.

In the selected EvoX design, we find that the seed's L2/LLC placement distinction and prefetch metadata are maintained, but both prediction mechanisms are replaced. Its L2 prefetcher learns, for each load instruction and within-page address delta, which nearby blocks are likely to be accessed soon afterward, placing high-confidence predictions in the L2 and lower-confidence predictions in the LLC. The confidence threshold increases with memory bandwidth utilization, unused prefetches, and cache pollution. The prefetcher additionally exploits previously observed page footprints when a page is revisited. The LLC replacement policy predicts near-term block reuse, temporarily protects likely reusable blocks, and uses page-level reuse history to control protection on insertion. The resulting design uses 127.64~KB of the 128~KB L2 prefetcher budget and 47.92~KB of the 48~KB LLC replacement budget.

\begin{figure}[t]
  \centering
  \includegraphics[width=\columnwidth]{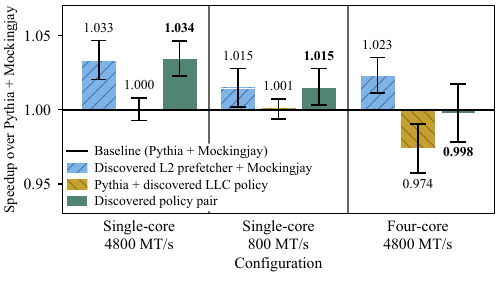}
  \caption{Policy ablation on held-out workloads, relative to baseline, across single-core and four-core configurations.}
  \label{fig:multicore}
\end{figure}

\subsubsection{\textbf{RQ1: Novelty.}}
We assess novelty relative to both the seed and prior research. A design is novel relative to the seed if it replaces core prediction mechanisms rather than merely tuning them. Against the 177-paper idea library, we distinguish mechanisms without published precedents from adaptations that extend prior work through changes in learned information, state maintenance, or architectural decisions.

\noindent\textbf{Novelty Assessment: L2 Prefetcher.} The EvoX champion replaces both prediction mechanisms of the seed. Its architectural specification references 16 papers from the idea library and identifies the mechanisms adapted from each. Although the prefetcher's individual components have published precedents, no paper in the library combines them into the same mechanism. Three adaptations extend prior work. First, each learned pattern adjusts its prefetching threshold based on both the usefulness and cache pollution of its own predictions, whereas prior designs maintain feedback globally or use pollution to disable prefetching for an instruction~\cite{srinath2007fdp,friendlyfire}. Second, previously observed page footprints influence the prediction threshold, rather than directly generating prefetch requests as in earlier footprint-based designs~\cite{bingo}. Third, the prefetcher learns nearby blocks accessed after each trigger, extending approaches that learn deltas between accesses of the same instruction~\cite{berti}.

\noindent\textbf{Novelty Assessment: LLC Replacement Policy.} The EvoX champion also replaces the seed's insertion-priority mechanism with prediction based protection of blocks likely to be reused. It learns reuse behavior from sampled cache sets, drawing on mechanisms used by Hawkeye and Mockingjay~\cite{jain2016hawkeye,shah2022mockingjay}. It further adapts page-level reuse counting, previously used to exclude blocks from caching~\cite{runtime_adpt}, to determine whether newly inserted blocks receive temporary protection. Thus, the replacement policy extends an existing mechanism to a different architectural decision, although its standalone performance remains comparable to Mockingjay, as shown by the policy ablation in Figure~\ref{fig:multicore} and discussed further in RQ2.

\begin{tcolorbox}[
    colback=insightbg,
    colframe=black,
    coltext=black,
    boxrule=0.6pt,
    arc=5pt,
    left=6pt,
    right=6pt,
    top=4pt,
    bottom=4pt,
    before skip=5pt,
    after skip=5pt
]
\noindent\textbf{Insight RQ1.}
The EvoX champion design replaces both core prediction mechanisms of its seed. Three architectural adaptations extend prior mechanisms, and no paper in the 177-paper idea library combines these mechanisms in the same design.
\end{tcolorbox}

\begin{figure}[t]
  \centering
  \includegraphics[width=\columnwidth]{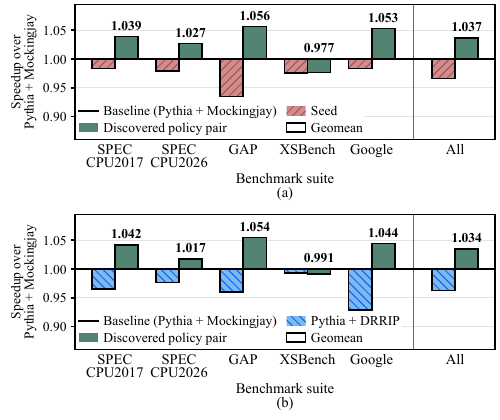}
  \caption{Speedup of the discovered policy pair w.r.t. baseline in single-core configuration. (a) On training workloads (30M warm-up and 50M measured instructions). (b) On held-out workloads (50M warm-up and 200M measured instructions).}
  \label{fig:speedup}
\end{figure}

\subsubsection{\textbf{RQ2: Evolution Efficacy.}}
On the training traces, the initial seed underperforms Pythia-Mockingjay by 3.4\%. \ourtool{} surpasses this baseline by candidate 5 and discovers the EvoX champion at candidate 38, achieving a GM speedup of 1.037. The 40-candidate search takes 53 hours (2.2 days), with the champion discovered after 51 hours. Of these candidates, 39 are successfully simulated and scored, while one is rejected before simulation (RQ4). Figure~\ref{fig:speedup} compares the seed and EvoX champion on training traces and the champion with Pythia-DRRIP on held-out traces, with all results normalized to Pythia-Mockingjay. On held-out traces, the champion achieves 3.4\% and 1.5\% higher performance at 4800 and 800~MT/s, respectively, compared with Pythia-Mockingjay, and 7.4\% and 4.6\% over Pythia-DRRIP. The seed is 3.8\% slower than Pythia-Mockingjay on held-out traces at 4800~MT/s, demonstrating a 7.2-percentage-point improvement from seed to champion.

\noindent\textbf{Policy Contributions and Multicore Performance.} Figure~\ref{fig:multicore} isolates the contributions of the evolved L2 prefetcher and LLC replacement policy. On single-core held-out workloads at 4800~MT/s, the evolved L2 prefetcher paired with Mockingjay achieves 3.3\% higher performance than Pythia-Mockingjay, while the evolved LLC replacement policy paired with Pythia matches the baseline. The jointly evolved pair achieves a speedup of 1.034, exceeding the product of the individual results by approximately 0.1\%, indicating that most of the single-core improvement comes from the evolved L2 prefetcher, with limited additional benefit from policy interaction. In the four-core configuration, where four cores share the LLC and one memory channel, the champion's harmonic speedup across 100 mixes of held-out traces is 0.2\% below Pythia-Mockingjay. However, the evolved L2 prefetcher, paired with Mockingjay, improves performance by 2.3\%, indicating that its benefit persists in multicore workloads, whereas the evolved LLC replacement policy offsets this gain. As the replacement policy was evolved for a single-core LLC, we manually scale its per-set state and two sampler tables by a factor of four for the multicore configuration, following Mockingjay~\cite{shah2022mockingjay}.

\begin{table}[t]
\centering
\caption{Performance comparison of \ourtool{} to other methodologies, across different LLMs and search strategies.}
\label{tab:results}
\scriptsize
\setlength{\tabcolsep}{2pt}
\renewcommand{\arraystretch}{0.95}
\setlength{\fboxsep}{1pt}
\resizebox{\columnwidth}{!}{%
\begin{tabular*}{\columnwidth}{@{\extracolsep{\fill}}llrrr@{}}
\toprule
\textbf{Model} &
\textbf{Methodology} &
\hspace{-3pt}\textbf{Train IPC} &
\hspace{-3pt}\textbf{Held-out IPC} &
\shortstack{\textbf{H/W Storage (KB)}\\\textbf{Prefetch / LLC}} \\
\midrule
\multirow{6}{*}{\shortstack[l]{OpenAI\\(GPT-6 Astra~\cite{openai2026gpt6astra})}}
& ArchAgent v1~\cite{gupta2026archagent}        & \textbf{1.004} & \textbf{1.000} & -- / 46.06 \\
& ArchAgent v2~\cite{gonzalez2026archagentv2}        & 1.000 & \textbf{1.000} & 25.00 / -- \\
& \ourtool-OpenEvolve & 1.001 & 1.007 & 116.33 / 46.22 \\
& \ourtool-GEPA       & 0.986 & 0.983 & 127.97 / 47.50 \\
& \ourtool-AdaEvolve  & 0.989 & 0.999 & 122.34 / 46.06 \\
& \ourtool-EvoX       & 0.978 & 0.985 & 127.87 / 48.00 \\
\midrule
\multirow{6}{*}{\shortstack[l]{Anthropic~\\(Claude Opus 5.5~\cite{anthropic2026opus55})}}
& ArchAgent v1~\cite{gupta2026archagent}        & 1.003 & 1.000 & -- / 46.67 \\
& ArchAgent v2~\cite{gonzalez2026archagentv2}        & 1.000 & 1.000 & 25.00 / -- \\
& \ourtool-OpenEvolve & 1.020 & 1.012 & 98.67 / 47.91 \\
& \ourtool-GEPA       & \colorbox[HTML]{CDFECD}{\textbf{1.041}} & \colorbox[HTML]{CDFECD}{\textbf{1.036}} & 127.97 / 47.88 \\
& \ourtool-AdaEvolve  & \colorbox[HTML]{CDFECD}{\textbf{1.053}} & \colorbox[HTML]{CDFECD}{\textbf{1.048}} & 127.24 / 47.93 \\
& \ourtool-EvoX       & \colorbox[HTML]{CDFECD}{\textbf{1.037}} & \colorbox[HTML]{CDFECD}{\textbf{1.034}} & 127.64 / 47.92 \\
\bottomrule
\end{tabular*}%
}
\end{table}

{\noindent\textbf{Comparison with Prior Approaches.} 
Table~\ref{tab:results} compares \ourtool{} with ArchAgent v1~\cite{gupta2026archagent} and ArchAgent v2~\cite{gonzalez2026archagentv2} under the same LLMs. As the original ArchAgent implementations are not publicly available, we independently reproduce their approaches in our experimental setup using OpenEvolve. Under Claude Opus 5.5, \ourtool{} achieves held-out speedups of 1.012–1.048 across four search strategies, compared with 1.000 for the reproduced ArchAgent baselines. AdaEvolve achieves the highest single-core speedup (1.048), while the EvoX design analyzed in RQ1 achieves 1.034. Under GPT-6 Astra, however, specification-level evolution does not outperform code-level evolution, indicating sensitivity to the underlying LLM (RQ3). Furthermore, unlike ArchAgent v1 and v2, which evolve individual cache replacement and prefetching policies, \ourtool{} jointly evolves L2 prefetching and LLC replacement under explicit hardware constraints.

\begin{tcolorbox}[
    colback=insightbg,
    colframe=black,
    coltext=black,
    boxrule=0.6pt,
    arc=5pt,
    left=6pt,
    right=6pt,
    top=4pt,
    bottom=4pt,
    before skip=5pt,
    after skip=5pt
]
\noindent\textbf{Insight RQ2.}

\ourtool{} achieves up to 4.8\% higher held-out single-core IPC than Pythia–Mockingjay under Claude Opus 5.5. The analyzed EvoX design improves performance by 3.4\% and 1.5\% at full and restricted bandwidth, respectively, but shows no significant multicore gain.
\end{tcolorbox}

\subsubsection{\textbf{RQ3: Search Representation.}}
We compare specification-level and code-level evolution under matched seed, idea library, training/held-out workloads, hardware storage budgets, search strategies, LLMs, and 40 proposed candidates per run. The key difference is the candidate representation: specification-level evolution modifies natural-language architectural specifications, which are translated into C++ simulator implementations, whereas code-level evolution modifies C++ implementations directly.

\noindent\textbf{Design Quality.} 
Table~\ref{tab:rq3-representation} shows that, under Claude Opus 5.5, specification level evolution achieves a higher geometric-mean held-out speedup across search strategies than code-level evolution (1.032 vs. 1.016). Specification-level evolution outperforms code-level evolution in three of four matched strategies, with OpenEvolve favoring code-level evolution. Held-out speedups range from 1.012 to 1.048 for specification-level evolution and from 1.008 to 1.030 for code-level evolution. Under GPT-6 Astra, however, code-level evolution performs better across all four strategies, indicating that the benefits of specification-level evolution depend on the underlying LLM.

\begin{table}[t]
\centering
\caption{Performance comparison of specification vs. code-level evolution across different LLMs and search strategies.}
\label{tab:rq3-representation}
\scriptsize
\setlength{\tabcolsep}{2pt}
\renewcommand{\arraystretch}{0.95}
\setlength{\fboxsep}{1pt}
\resizebox{\columnwidth}{!}{%
\begin{tabular*}{\columnwidth}{@{\extracolsep{\fill}}lllrrrrr@{}}
\toprule
\textbf{Model} &
\textbf{Strategy} &
\textbf{Repr.} &
\shortstack{\textbf{Train}\\\textbf{IPC}} &
\shortstack{\textbf{Held-out}\\\textbf{IPC}} &
\shortstack{\textbf{LLM}\\\textbf{Calls}} &
\shortstack{\textbf{Tokens}\\\textbf{(M)}} &
\shortstack{\textbf{Cost}\\\textbf{(\$)}} \\
\midrule
\multirow{8}{*}{\shortstack[l]{OpenAI\\(GPT-6 Astra~\cite{openai2026gpt6astra})}}
& OpenEvolve~\cite{sharma2025openevolve} & Spec. & 1.001 & 1.007 & 267 & 1.22 & 228 \\
&            & C++   & \textbf{1.023} & \textbf{1.022} & 158 & 0.60 & 172 \\
\cmidrule(lr){2-8}
& GEPA~\cite{agrawal2025gepa}       & Spec. & 0.986 & 0.983 & 214 & 1.13 & 191 \\
&            & C++   & \textbf{1.024} & \textbf{1.021} & 162 & 0.73 & 189 \\
\cmidrule(lr){2-8}
& AdaEvolve~\cite{cemri2026adaevolve}  & Spec. & 0.989 & 0.999 & 273 & 1.55 & 253 \\
&            & C++   & \textbf{1.010} & \textbf{1.013} & 152 & 0.60 & 168 \\
\cmidrule(lr){2-8}
& EvoX~\cite{liu2026evox}       & Spec. & 0.978 & 0.985 & 275 & 1.43 & 257 \\
&            & C++   & \textbf{1.006} & \textbf{0.996} & 170 & 0.74 & 197 \\

\midrule
\multirow{8}{*}{\shortstack[l]{Anthropic\\(Claude Opus 5.5~\cite{anthropic2026opus55})}}
& OpenEvolve~\cite{sharma2025openevolve} & Spec. & 1.020 & 1.012 & 241 & 5.80 & 256 \\
&            & C++   & \colorbox[HTML]{CDFECD}{\textbf{1.040}} & \textbf{1.030} & 168 & 2.71 & 167 \\
\cmidrule(lr){2-8}
& GEPA~\cite{agrawal2025gepa}       & Spec. & \colorbox[HTML]{CDFECD}{\textbf{1.041}} & \colorbox[HTML]{CDFECD}{\textbf{1.036}} & 224 & 5.44 & 248 \\
&            & C++   & 1.020 & 1.012 & 146 & 2.13 & 146 \\
\cmidrule(lr){2-8}
& AdaEvolve~\cite{cemri2026adaevolve}  & Spec. & \colorbox[HTML]{CDFECD}{\textbf{1.053}} & \colorbox[HTML]{CDFECD}{\textbf{1.048}} & 203 & 3.95 & 184 \\
&            & C++   & 1.022 & 1.008 & 168 & 2.71 & 169 \\
\cmidrule(lr){2-8}
& EvoX~\cite{liu2026evox}       & Spec. & \textbf{1.037} & \colorbox[HTML]{CDFECD}{\textbf{1.034}} & 274 & 7.21 & 320 \\
&            & C++   & 1.029 & 1.015 & 163 & 2.65 & 169 \\
\bottomrule
\end{tabular*}%
}
\end{table}

\noindent\textbf{Search Efficiency.} Table~\ref{tab:candidate-outcomes} shows that specification-level evolution successfully scores 245 of 320 proposed candidates (76.6\%), compared with 292 of 320 (91.3\%) for code-level evolution. Only one specification-level candidate fails specification validation and eight fail build-stage checks, while most unscored candidates stop at the simulation stage (66 candidates). This lower throughput reflects the additional specification generation and validation steps, along with LLM rate limits and execution-time constraints, which also account for some candidates classified as simulation-stage failures.

\noindent\textbf{Computational Cost.} As shown in Table~\ref{tab:rq3-representation}, specification-level evolution incurs 1.55$\times$ greater estimated LLM cost under Claude Opus 5.5 (\$1,008 vs.\ \$651) and 1.28$\times$ under GPT-6 Astra (\$929 vs.\ \$726) across four search strategies. It also requires more LLM calls and approximately twice as many output tokens. However, ChampSim evaluation times remain comparable across representations (3,200--4,800 seconds per candidate), indicating that the additional overhead arises primarily from architectural specification generation and validation rather than simulation.

\begin{tcolorbox}[
    colback=insightbg,
    colframe=black,
    coltext=black,
    boxrule=0.6pt,
    arc=5pt,
    left=6pt,
    right=6pt,
    top=4pt,
    bottom=4pt,
    before skip=5pt,
    after skip=5pt
]
\noindent\textbf{Insight RQ3.} 

Under Claude Opus 5.5, specification-level evolution achieves higher held-out performance than code-level evolution (3.2\% vs. 1.6\%) across four strategies, at 1.55× greater LLM cost.

\end{tcolorbox}

\subsubsection{\textbf{RQ4: Candidate Validation.}}
In the selected Claude Opus 5.5/EvoX specification-level run, \specgate{} returned five specification versions due to a storage-table entry that could not be accounted for, while \checkf{} and \checkd{} flagged nine blocking issues across eight specifications. These issues included arithmetic overflow, conditions that could halt L2 prefetching, a policy requiring per-cycle actions unsupported by the replacement interface, and a policy reproducing a published design. All 40 specifications passed initial validation, including eight after 1-2 revisions. However, one candidate was rejected before simulation after a \codewriter{} correction caused its specification to fail a consistency re-check. Table~\ref{tab:candidate-outcomes} summarizes the final outcomes across all runs. In the selected Claude Opus 5.5/EvoX run, all 39 specification-level candidates that reached simulation completed successfully, while all 40 code-level candidates passed the corresponding C++ model checks and simulation. Other LLM and search-strategy combinations experienced simulation-stage failures, indicating that this result does not generalize to all specification-level runs.

\begin{tcolorbox}[
    colback=insightbg,
    colframe=black,
    coltext=black,
    boxrule=0.6pt,
    arc=5pt,
    left=6pt,
    right=6pt,
    top=4pt,
    bottom=4pt,
    before skip=5pt,
    after skip=5pt
]

\noindent\textbf{Insight RQ4.} In the Claude Opus 5.5/EvoX run, staged validation revised 8 candidates and rejected 1 before simulation; all 39 admitted candidates simulated successfully.

\end{tcolorbox}

\begin{table}[t]
\centering
\caption{Candidate outcomes across different LLMs and search strategies for the final stage of candidates.}
\label{tab:candidate-outcomes}
\scriptsize
\setlength{\tabcolsep}{2pt}
\renewcommand{\arraystretch}{0.95}
\resizebox{\columnwidth}{!}{%
\begin{tabular*}{\columnwidth}{@{\extracolsep{\fill}}lllrrrrl@{}}
\toprule
& & & &
\multicolumn{3}{c}{\textbf{Final stage}} & \\
\cmidrule(lr){5-7}
\textbf{Model} &
\textbf{Strategy} &
\textbf{Repr.} &
\textbf{Proposed} &
\textbf{Spec.} &
\textbf{Build} &
\textbf{Sim.} &
\textbf{Scored} \\
\midrule
\multirow{8}{*}{\shortstack[l]{Anthropic\\(Claude Opus 5.5~\cite{anthropic2026opus55})}}
& OpenEvolve~\cite{sharma2025openevolve} & Spec. & 40 & 0 & 1 & 10 & 29 (72.5\%) \\
&            & C++   & 40 & NA & 0 & 3 & 37 (92.5\%) \\
\cmidrule(lr){2-8}
& GEPA~\cite{agrawal2025gepa}       & Spec. & 40 & 0 & 0 & 9 & 31 (77.5\%) \\
&            & C++   & 40 & NA & 0 & 8 & 32 (80.0\%) \\
\cmidrule(lr){2-8}
& AdaEvolve~\cite{cemri2026adaevolve}  & Spec. & 40 & 1 & 1 & 12 & 26 (65.0\%) \\
&            & C++   & 40 & NA & 0 & 4 & 36 (90.0\%) \\
\cmidrule(lr){2-8}
& EvoX~\cite{liu2026evox}       & Spec. & 40 & 0 & 1 & 0 & 39 (97.5\%) \\
&            & C++   & 40 & NA & 0 & 0 & 40 (100\%) \\
\midrule
\multirow{8}{*}{\shortstack[l]{OpenAI\\(GPT-6 Astra~\cite{openai2026gpt6astra})}}
& OpenEvolve~\cite{sharma2025openevolve} & Spec. & 40 & 0 & 1 & 9 & 30 (75.0\%) \\
&            & C++   & 40 & NA & 0 & 1 & 39 (97.5\%) \\
\cmidrule(lr){2-8}
& GEPA~\cite{agrawal2025gepa}       & Spec. & 40 & 0 & 3 & 15 & 22 (55.0\%) \\
&            & C++   & 40 & NA & 2 & 3 & 35 (87.5\%) \\
\cmidrule(lr){2-8}
& AdaEvolve~\cite{cemri2026adaevolve}  & Spec. & 40 & 0 & 1 & 9 & 30 (75.0\%) \\
&            & C++   & 40 & NA & 0 & 7 & 33 (82.5\%) \\
\cmidrule(lr){2-8}
& EvoX~\cite{liu2026evox}       & Spec. & 40 & 0 & 0 & 2 & 38 (95.0\%) \\
&            & C++   & 40 & NA & 0 & 0 & 40 (100\%) \\
\bottomrule
\end{tabular*}%
}
\end{table}
\section{Discussion}
\label{sec:discussion}

\noindent\textbf{Research-derived architectural novelty.}
A central capability of \ourtool{} is its use of architectural ideas extracted from published research, rather than restricting candidate generation to modifications of an existing implementation. The discovered design illustrates how ideas with precedents in prior work can be adapted into mechanisms that differ in their learned information, state organization, or decisions. However, the novelty assessment also shows that the individual mechanisms are not necessarily unprecedented. The contribution of research-guided generation is therefore the ability to explore new uses and combinations of architectural ideas, not a guarantee that every generated component is fundamentally new outside of its idea library.

\noindent\textbf{Validation and hardware realizability.}
Evolving architectural specifications introduces additional validation requirements and computational overhead. Although \ourtool{} checks architectural feasibility, specification–implementation consistency, and hardware storage constraints, these checks do not guarantee correctness or hardware realizability. In particular, storage accounting does not constrain combinational logic complexity or execution latency. ChampSim executes policy logic within a simulator event without modeling its hardware latency, potentially favoring complex mechanisms that would require multiple cycles in hardware. Such delays could reduce prefetch timeliness and diminish the performance gains observed in simulation.

\noindent\textbf{Generalizability and limitations.} Our evaluation focuses on LLC replacement and L2 prefetching as a case study of interacting microarchitectural policies. Although \ourtool{} jointly represents their mechanisms and permitted information exchange, the communication interface remains fixed, limiting exploration of interactions requiring changes to other simulator components. The discovered design improves single-core performance but shows no statistically significant gain over Pythia--Mockingjay in the evaluated multicore configuration, potentially because single-core-guided evolution does not capture multicore resource contention. Moreover, the methodology depends on the coverage and fidelity of the research-derived idea library, LLM capabilities, and architectural interfaces. Extending it to other policy types, communication interfaces, and multicore-aware search remains future work.
\section{Conclusion}
\label{sec:conclusion}

In this work, we present \ourtool{}, an LLM-guided methodology for generating microarchitectural policies through evolving natural-language architecture specifications, rather than simulator code. \ourtool{} first leverages ideas from prior microarchitectural studies through extracting and organizing concepts from prior works into a library. Staged validation then enforces preliminary checks for hardware constraints and consistency between specifications and generated C++ simulation code, enabling simulation-derived performance rewards and diagnostic feedback to inform subsequent architectural revisions. Through the joint evolution of LLC replacement and L2 prefetching policies, \ourtool{} demonstrates the effectiveness of research-guided specification-level exploration under hardware constraints. The discovered design improves geometric-mean IPC on held-out workloads by 3.4\% and 1.5\% over Pythia-Mockingjay at full and restricted memory bandwidth, respectively, and by 7.4\% and 4.6\% over the Pythia--DRRIP baseline. Starting from a seed design that underperforms Pythia--Mockingjay, \ourtool{} achieves a training speedup of 3.7\% within 38 candidate evaluations. These results demonstrate the potential of evolving explicit architectural specifications to discover effective microarchitectural mechanisms beyond direct refinement of existing simulator codes.

\section{Acknowledgments}
\label{sec:ack}
All research, experimental design, implementation, data analysis, and reported results were performed independently by the authors. The use of artificial intelligence (AI) tools was strictly limited to assisting with manuscript preparation and editorial refinement. The authors acknowledge the support from the Center for Secure Microelectronics Ecosystem — CSME\#210205.

\bibliographystyle{ACM-Reference-Format}
\bibliography{refs}

\end{document}